\documentclass[pdflatex,sn-mathphys-num]{sn-jnl}

\usepackage{graphicx}%
\usepackage{multirow}%
\usepackage{amsmath,amssymb,amsfonts}%
\usepackage{amsthm}%
\usepackage{mathrsfs}%
\usepackage[title]{appendix}%
\usepackage{xcolor}%
\usepackage{textcomp}%
\usepackage{manyfoot}%
\usepackage{booktabs}%
\usepackage{algorithm}%
\usepackage{algorithmicx}%
\usepackage{algpseudocode}%
\usepackage{listings}%

\theoremstyle{thmstyleone}%
\theoremstyle{thmstyletwo}%

\theoremstyle{thmstylethree}%

\begin{document}

\title[Article Title]{The time-delay model and its applications to galactic archaeology}


\author*[1,2,3,4]{\fnm{Francesca} \sur{Matteucci}}\email{francesca.matteucci@inaf.it}



\affil*[1]{\orgdiv{Department of Physics}, \orgname{Trieste University}, \orgaddress{\street{Via Fabio Valerio 2}, \city{Trieste}, \postcode{34143}, \state{Friuli-Venezia-Giulia}, \country{Italy}}}

\affil[2]{\orgdiv{Osservatorio Astronomico di Trieste}, \orgname{INAF}, \orgaddress{\street{Via G.B. Tiepolo, 11}, \city{Trieste}, \postcode{34131}, \state{Friuli-Venezia-Giulia}, \country{Italy}}}

\affil[3]{\orgdiv{Sezione di Trieste}, \orgname{INFN}, \orgaddress{\street{Via Fabio Valerio, 2}, \city{Trieste}, \postcode{34143}, \state{Friuli-Venezia-Giulia}, \country{Italy}}}

\affil[4]{\orgname{IFPU, Institute for the Fundamental Physics of the Universe}, \orgaddress{\street{Via Beirut, 2}, \city{Trieste}, \postcode{34151}, \state{Friuli-Venezia-Giulia}, \country{Italy}}}


\abstract{The time-delay model is the way we interpret the diagram [X/Fe] vs. [Fe/H], where $X$ is the abundance of a generic element from carbon to uranium. This interpretation is based on the lifetimes of stars of different masses producing different elements. The abundance of Fe ([Fe/H]) traces  the "stellar metallicity" and  is due to supernovae Type Ia, which are believed to be the major producers of Fe, and in part to supernovae core-collapse. In particular, if $X$ is an $\alpha$-element, produced on short timescales from massive stars, the ratio [$\alpha$/Fe] will show an overabundance of the $\alpha$-elements relative to Fe at low metallicity. In fact, the bulk of Fe is produced with a time delay relative to $\alpha$-elements, since Type Ia supernovae are white dwarfs in binary systems and they can have lifetimes as long as the age of the Universe. In this paper, I will show how powerful is the time-delay model in order to interpret the abundance patterns observed in stars and interstellar gas, since it allows us to put constraints on stellar nucleosynthesis as well as on the star formation histories of galaxies. I will present some applications of the time-delay model, in particular to the chemical evolution of the Milky Way  and galaxies of different morphological type as well as to the identification of high redshift objects by means of their abundances.}


\keywords{chemical abundances, galaxy evolution, stellar nucleosynthesis,}



\maketitle

\section{Introduction}\label{sec1}

Galactic archaeology consists in reconstructing the star formation history of a galaxy by means of the interpretation of their abundance patterns. The most studied galaxy is the Milky Way (MW), for which we have now millions of stellar data.  The field of galactic chemical evolution was started by \citep{tinsley1980}. To build a model for the chemical evolution of the MW we need some necessary ingredients. They are: i) initial conditions, ii) stellar nucleosynthesis (stellar yields), iii) birthrate function, namely the product of the star formation rate (SFR)  and  initial mass function (IMF), gas flows (accretion and galactic winds). Chemical evolution models predict the evolution in time and space of the interstellar medium (ISM) and its chemical composition in a given galaxy. The predicted abundances of chemical elements are then compared either with the abundances measured in the atmospheres of stars of different ages and metal content or with the abundances in the ISM. From this comparison we can derive constraints on stellar yields as well as on the star formation history and gas flows. 
The main parameters in such models are therefore the stellar yields, SFR, IMF and gas flows in and out. 
For the star formation law, there are observations \citep{kennicutt1989, kennicutt1998, schmidt1963} suggesting that the SFR is proportional to the gas density. Concerning the IMF, it has been derived only in the solar vicinity by stellar counts and the first study was that of \citep{salpeter1955}, followed by other studies  (\citep{scalo1986, arimoto1987, kroupa1993, chabrier2003}) all indicating that the IMF is a power law. For the gas flows, one can assume that galactic disks are formed by gas accretion that in turn triggers radial gas flows. Galactic winds can also be present and occur when the thermal energy of the gas is larger than its binding energy. 
Many models of galactic chemical evolution have been developed after \cite{tinsley1980} and I will mention only a few, in particular those which introduced particular scenarios for the MW formation, such as the one-infall model of \cite{matteucci1986},  the inside-out disk formation models of \cite{matteucci1989}  and \cite{boissier1999}, the parallel model of \cite{pardi1995}, the two-infall model of \cite{chiappini1997}, the inhomogeneous halo models of \cite{argast2002} and \cite{cescutti2008}. For more references see \cite{matteucci2021}.

The concept of "time-delay" model was first introduced by \cite{tinsley1980}  and then by \citep{greggio1983} who computed the Type Ia SN rate in the framework of the single-degenerate scenario, where the system  is formed by a C-O white dwarf plus a normal star. In \citep{matteucci1986} the formulation of the Type Ia SN rate of \citep{greggio1983} was introduced for the first time in a detailed chemical evolution model. This allowed them to interpret the       [O/Fe] vs. [Fe/H] diagram as due to the delayed Fe production relative to O. In that paper, it was also predicted that other $\alpha$-elements such as Mg and Si should follow the same trend, and \cite{francois1988} data confirmed that this suggestion was indeed correct. In the paper of \citep{matteucci1986} it was also suggested that the timescale of formation of the stellar halo should have been roughly 1 Gyr and this time corresponds to [Fe/H] $\sim$ -1.0 dex, the value at which the [O/Fe] ratio shows a knee, namely a rapid change of slope due to the Fe that starts to be produced in a substantial way by Type Ia SNe.
It is worth noting that the very first Type Ia SNe in the model of \citep{greggio1983} explode already after 35 Myr, which is the lifetime of a $8M_{\odot}$ star, but their contribution to Fe production becomes important only after 1Gyr. This finding let to think that 1 Gyr was a universal value marking the occurrence of Type Ia SNe, but this is not correct. In fact, in \cite{matteucci1990} it was predicted that the relation [O/Fe] vs. [Fe/H] in the stars of the Galactic bulge is different than that in the solar vicinity, with a knee  at a larger metallicity, namely at [Fe/H]$\sim$0 dex, and this was later confirmed by observations (e.g. \citep{mcwilliam1994}). On the other hand, in dwarf irregular galaxies the knee should occur at metallicities much lower than [Fe/H]=-1.0 dex. 
In the following years, the time-delay model has become the best interpretation for the diagram [X/Fe] vs. [Fe/H] and it can be applied to any element with abundance $X$.\\
The paper is organized as follows: in Section 2 we show the first attempts to explain the abundance patterns in the solar vicinity and discuss the constraints that can be derived; in Section 3 we present the results obtained for external galaxies of different morphological type and
show how we can infer the morphology of high-redshift galaxies just by means of chemical abundances. We also describe how to identify the galaxy type of Gamma Ray Burst (GRB) hosts.
Finally, in Section 4 we present a discussion and conclusions.

\section{Chemical evolution models\label{sec2}}
The basic ingredients of chemical evolution models are:
\begin{itemize}
\item {\bf Initial conditions}: gas present since the beginning or accreted subsequently. Gas with primordial chemical composition or pre-enriched by Population III stars;
\item {\bf The history of star formation}: the birthrate function can be written as:
\begin{equation}
B(m,t) =\psi(t) \varphi(m)dm dt
\end{equation}
depending upon  stellar mass and time. 
The most common parametrization of the SFR ($\psi(t)$)is a power law depending upon the gas density (Kennicutt-Schmidt law):
\begin{equation}
    \psi(t)=\nu \sigma(t)_{gas}^k
\end{equation}
  where k=1.5$\pm$0.15 (\cite{kennicutt1998}) and $\nu$ is the efficiency of star formation.  \\
  
  The IMF ($\varphi(m)$)is also a power law of the type:
  \begin{equation}
      \varphi(m) \propto m^{(1+x)}
  \end{equation}
where $x$ is known as Salpeter index. The IMF has been derived only for solar neighborhood stars and described with more than one slope (\citep{miller1979, tinsley1980, scalo1986, kroupa2001, chabrier2003, weidner2005}).\\
The rates of SNe of all types (II, Ia, Ib, Ic) are computed in detail: in particular, for Type Ia SNe, originating from white dwarfs in binary systems, we adopt a Delay Time Distribution  (DTD) function describing the explosion times of these SNe. The DTD can be related to theoretical models, such as the single degenerate (SD) or the double degenerate (DD) one, but also empirically derived. In our models we adopt the  DTD of the SD scenario,  which is very similar to that of the DD one, since it reproduces at best the abundances in galaxies (\cite{matteucci2009}) as well as the cosmic Type Ia SN rate (see \cite{palicio2024}). The empirical DTD $\propto t^{-1}$ (\cite{totani2008}) can also reasonably reproduce the observations but is less precise at early times.
In order to compute the Type Ia SN rate and the rate of ejection from these objects, we multiply the assumed DTD by the SFR 
(see \cite{matteucci2021} for a review and details).

\item {\bf Stellar yields}: stellar nucleosynthesis is a fundamental parameter in chemical evolution models, where we should take into account both the newly formed and already present elements. 
Stars of different masses produce different chemical species: massive stars ($M> 8M_{\odot}$) are responsible for the formation of $\alpha$-elements, r-process elements, some Fe and some light s-process elements. Low and intermediate mass stars (LIMS) ($0.8< M/M_{\odot}<8$) are responsible for the production of He, $^{14}N$,  $^{12}C$ and $^{13}C$ , heavy s-process elements. Type Ia SNe belong to the same mass range of LIMS, since they originate from C-O white dwarfs: they produce the bulk of Fe. 
Massive stars explode as core-collapse SNe (CC-SNe) and they can be Type II or Type Ib,c, according to their initial mass. 
Concerning r-process elements, they are predicted to be produced by rare classes of CC-SNe (e.g., magneto-rotational supernovae and/or collapsars; see \cite{mosta2018} and \cite{siegel2019}. The only confirmed scenario is, however, the merging of two neutron stars (see [22]), as shown by the kilonova AT2'17gfo, following the gravitational wave event GW170817 (see [23]).
The stellar yields have been computed by several authors, I will mention here for massive stars:  \citep{woosley1995}, \citep{nomoto2013}, \citep{limongi2018}. For LIMS: \citep{marigo2001},\citep{karakas2010}, \citep{ventura2013}, among others.
For Type Ia SNe: \citep{iwamoto1999}, \citep{leung2018}.\\

\item {\bf Gas flows}: infall, outflow and radial gas flows are fundamental processes to shape the dynamics and chemistry of a galaxy.\\
  The infall rate can be expressed by means of an exponential law:
  \begin{equation}
      I(t)=ae^{-t/\tau}
  \end{equation}
  where $a$ is a parameter fixed by reproducing the total surface mass density at the present time, and $\tau$ is the time scale for gas accretion.\\
  
  The wind can be expressed as proportional to the SFR, since stellar feedback is a fundamental source of energy for the ISM:
  \begin{equation}
      W(t)=\lambda \cdot \psi(t)
  \end{equation}
  with $\lambda$ being the mass loading factor which is treated as a free parameter.
\end{itemize}
 All of these ingredients are then included in integer-differential equations, one for each element, from H to uranium. In particular:
\begin{equation}
    \sigma_{gi}(t)=-X_i(t) \psi(t) + R_i(t) + I_i(t) - W_i(t),
\end{equation}

where $\sigma_{gi}(t)$ is the surface gas density in the form of the generic element $i$ and $X_i(t)$ is the abundance  by mass of the element $i$. The quantity $R_i(t)$ is the rate of restitution to the ISM of the element $i$ by dying stars, and contains the stellar nucleosynthesis and IMF, while $I_i(t)$ and $W_i(t)$ are the rate of infall and outflow of the element $i$, respectively.
By solving this equation we can predict the time evolution of the abundance of a given element present in the ISM.

\subsection{Interpretation of abundance patterns in the solar vicinity: time-delay model}\label{sec3}
In Figure 1, the predictions for [O/Fe] vs. [Fe/H] from \cite{matteucci1986} models are presented and compared to the available data at that time. This Figure illustrates clearly the time-delay model: the line labelled $b$ represents a model where all the Fe is (incorrectly) assumed to be produced by Type Ia SNe,
while O is assumed (correctly) to originate only from massive stars; the results 
clearly do not fit the data, since the [O/Fe] is continuously decreasing. On the other hand, model labeled $a$ assumes (correctly) that 70\% of Fe is produced by Type Ia SNe and the remaining 30\% by CC-SNe. Clearly, to reproduce the plateau of [O/Fe] at low metallicities and the following decrease for [Fe/H]$>$-1.0 dex, we need to assume that while O originates only from massive stars, while Fe is produced mainly in Type Ia SNe but also in CC-SNe. The knee, where the [O/Fe] ratio starts decreasing, marks the end of the halo formation phase and corresponds, in Model $a$, to roughly an age of 1Gyr. This timescale was then interpreted by the astronomical community, as a fundamental constant valid in every galaxy, but this is not the case. Later on, \cite{matteucci1990} showed that for the bulge or a spheroid such an elliptical galaxy, the knee should occur at larger metallicities, while in dwarf galaxies it should appear at lower metallicities. As a consequence of this, we could have low [$\alpha$/Fe] ratios at low [Fe/H] and high [$\alpha$/Fe] at high [Fe/H]. This variety of situations is explained by considering that galaxies of different morphological type have different star formation histories (SFHs), being the SFR very high in spheroids and very low in dwarf galaxies, with the spirals like the MW being in the middle. 

\begin{figure}[h]
\centering
\includegraphics[width=0.7\textwidth]{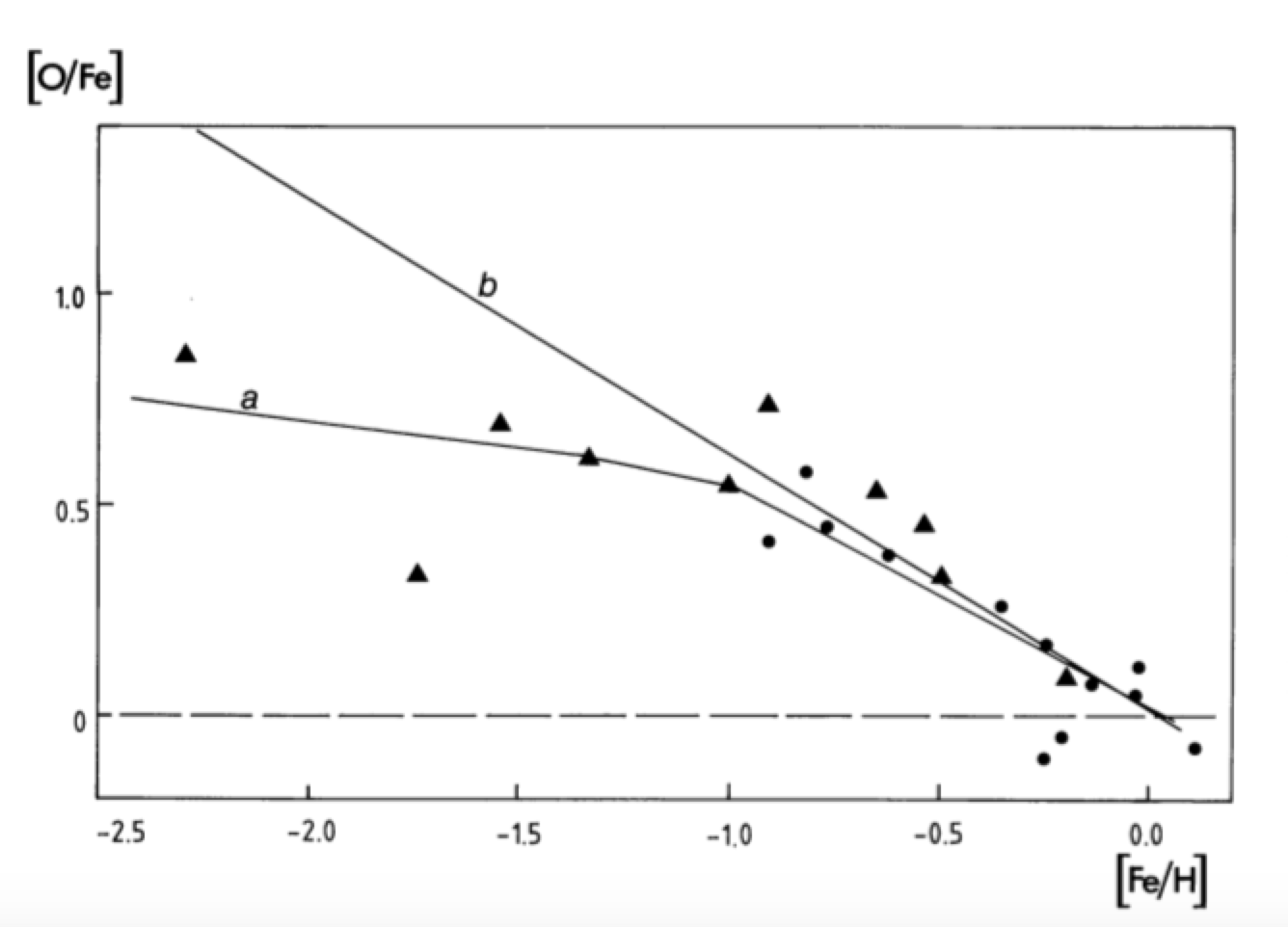}
\caption{The [O/Fe] vs. [Fe/H] diagram for the solar vicinity. The models and data are from \cite{matteucci1986}. The models are described in the text.}
\label{fig1}
\end{figure}

In the following years, the plot [X/Fe] vs.[Fe/H] was extended to all the chemical species including C,N,O, $\alpha$-elements, Fe, Fe-peak elements, s-and r-process elements. One example is shown in Figure 2, where the predictions of the two-infall model of \citep{romano2010} are compared to data. The models plotted in Figure 2 refer to two different sets of stellar yields for massive stars. What is interesting to note is that if a generic element $X$ is produced in the same proportions and by the same stars as Fe, then the [X/Fe] ratio appears constant and solar along the whole metallicity range. On the other hand, secondary elements (produced from metals already present in the star) show an increase of the [X/Fe] ratio with metallicity in the range of low [Fe/H] values, and this is because the abundance of such elements is proportional to metallicity. It should be said that N, in that plot,  is treated as a secondary element from massive stars except for those extremely metal poor, and that this is not the case since data show a flat [N/Fe] behavior at low metallicity and this is explained only if massive stars produce primary N (\citep{matteucci1986}) and this is indeed predicted by models of rotating massive stars (e.g.\citep{meynet2002, limongi2018}). It is worth noting that in Figure 2 the fit is bad for certain elements, among which the already mentioned N, K and some Fe-peak elements. This is due to the large uncertainties still present in stellar nucleosynthesis.\\
Other important chemical species, whose nature is still uncertain are the s- and r- process elements. The difficulty with these n-rich elements is that they can be formed in part by slow neutron capture (s-process) and in part by rapid neutron capture (r-process). Elements with a predominant s-process component are Sr, Y, Zr (1rst peak), Ba, La, Ce (2nd peak), Pb (3rd peak), and are mainly produced in low mass stars (1-3$M_{\odot}$) during He-shell burning. As stated already in the Introduction, AGB stars are mainly responsible for the formation of the main and of the strong s-process, while rotating massive stars can produce the weak s-process.
A typical r-process element (97\%) is  Eu which should be produced during explosive nucleosynthesis in CC-SNe, although several authors (see for example \citep{arcones2011, thielemann2017}) have claimed that it is difficult to produce such elements in this way. 
On the other hand, merging neutron stars seem more promising sources of r-process elements (\citep{korobkin2012}). It has been shown by \citep{matteucci2014}  that merging neutron stars can produce all the Eu observed in the solar system under the strict conditions that all systems merge on a timescale no longer than 1 Myr and that all star up to $50M_{\odot}$ leave a neutron st<ar as a remnant. Other papers (\cite{simonetti2019, cote2019, molero2021, greggio2021}) have shown that one should assume a DTD function for the merging times and assume that at early times massive stars should be responsible for Eu production. This is because Eu behaves exactly as a $\alpha$-element in the plot [X/Fe] vs. [Fe/H], although a large spread is present in the data at low metallicity. In fact, according to the time-delay model, an element showing overabundance relative to Fe at low metallicity indicates that it is produced faster than Fe, namely mainly by massive stars.

\begin{figure}
\centering
\includegraphics[width=1.0\linewidth]{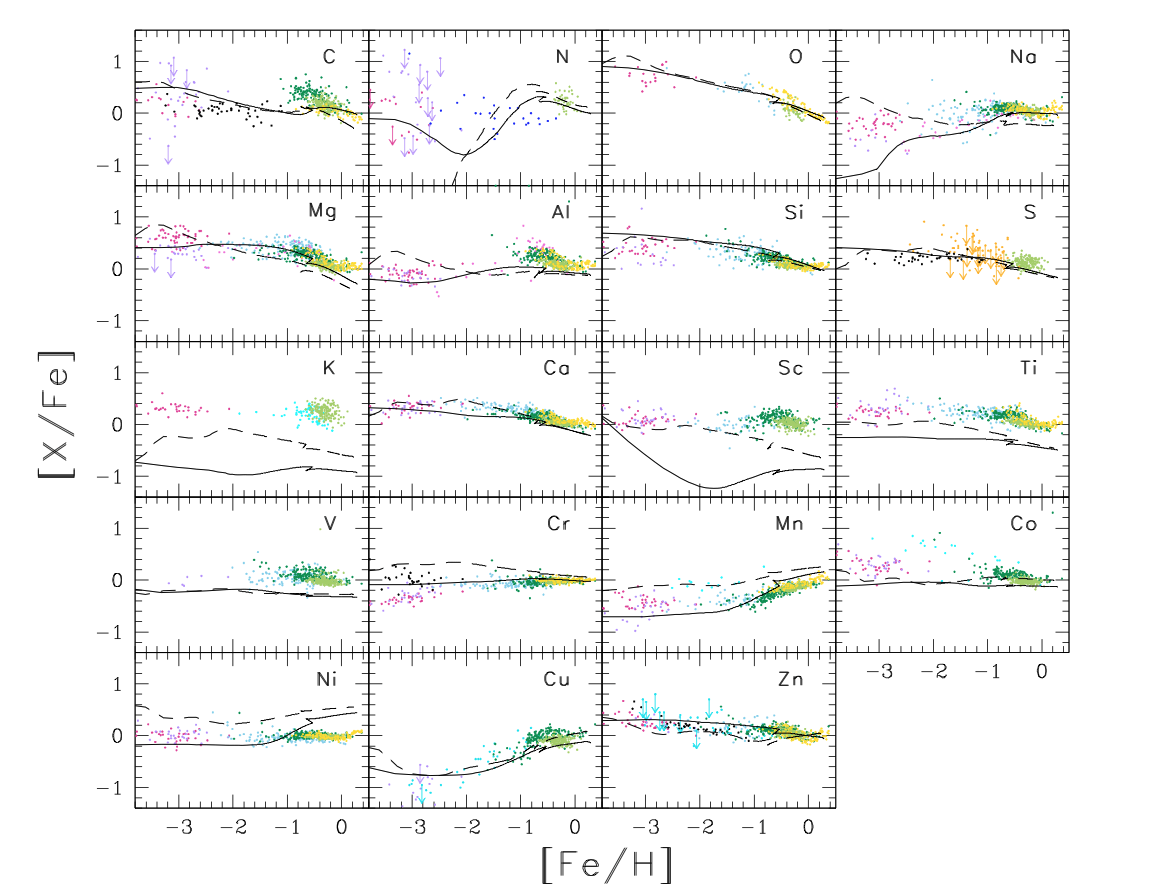}
\caption{The [X/Fe] vs. [Fe/H] plot for several chemical species, from carbon to zinc in the solar neighborhood. The model predictions are indicated by solid and dashed lines. They differ only for the adopted stellar yields (see \citep{romano2010}).}
    \label{fig2}
 \end{figure}

In Figure 3 we show the behavior of [Eu/Fe] vs. [Fe/H] in the solar neighborhood. 

\begin{figure}
\centering
\includegraphics[width=1.0\linewidth]{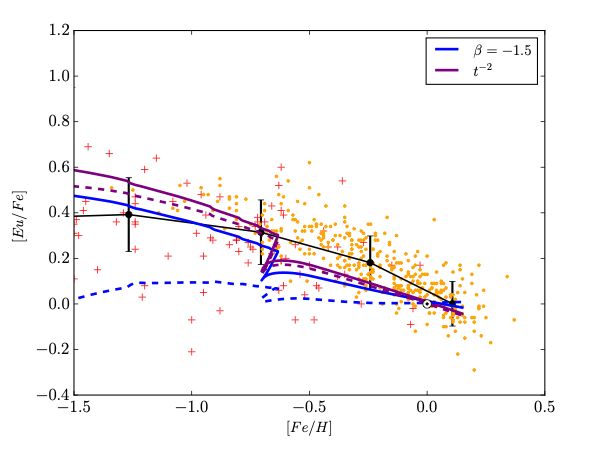}
\caption{The [Eu/Fe] vs. [Fe/H] plot in the solar neighborhood. The solid curve represents  models where Eu is co-produced by CC-SNe and merging neutron stars, while the dashed lines are models with only merging neutron stars. Blue and purple lines refer to different assumptions about the DTD function of merging neutron stars. Note that the DTD$\propto t^{-2}$ (probably unrealistic) predicts many systems merging at early times and therefore it fits much better the data.  Black points with error bars are the average observed [Eu/Fe] in a 0.5 dex wide bin. Figure from \citep{simonetti2019}.}
    \label{fig3}
 \end{figure}

Clearly, the Eu behaves like an $\alpha$-elements and needs fast producers, at least at early times, to reproduce the overabundance of Eu relative to Fe at low [Fe/H]. These producers can only be CC-SNe but there are many uncertainties on the r-process production during these events. An alternative explanation could be that,  given a DTD for merging neutron stars, necessary to explain the event GW170817 that occurred in an early type galaxy with negligible star formation and therefore indicating a long merging time, the fraction of binary neutron stars should have been higher at early times, as suggested by \citep{simonetti2019}.\\
Before closing this Section, I would like to remind two papers I co-authored with {\bf Roberto Gallino}: the first was on the origin of Cu and Zn by applying the time-delay model (\citep{matteucci1993}) and the second dealt with the  Si , Ti and Mg isotopic anomalies which were interpreted again by means of the time-delay model (\citep{gallino1994}). \\
It is worth showing more recent work (\cite{spitoni2024}) where the time-delay model is adopted to explain the apparent bimodality in the [$\alpha$/Fe] ratios in stars belonging to the thick and thin disks in the solar vicinity.  In Figure 4 we show a comparison between a revised version of the two-infall model applied to the thick and thin disks (\citep{spitoni2024}). In this framework, the thick and thin disks form out of two different gas accretion episodes, with the thick disk forming faster. Between the formation of the two disks there is a gap in the SFR, although it does not go to zero. This gap produces the loop shown in the Figure and is due to the fact that when star formation becomes negligible, the abundance of Fe decreases due to the dilution by the second infall, although Fe continues to be produced by Type Ia SNe, while the  $\alpha$/Fe ratio remain constant because the infall dilutes in the same way  all the elements. Then, when star formation resumes, the [$\alpha$/Fe] ratio starts decreasing again. In Figure 4 the various colors indicates three sequences: high (red), intermediate (green) and low (blue)  $\alpha$ stars and their percentages, compared to the observed ones, together with the few stars that are predicted to form in the period of the low star formation (indicated in pink). 

\begin{figure}
\centering
\includegraphics[width=0.9\linewidth]{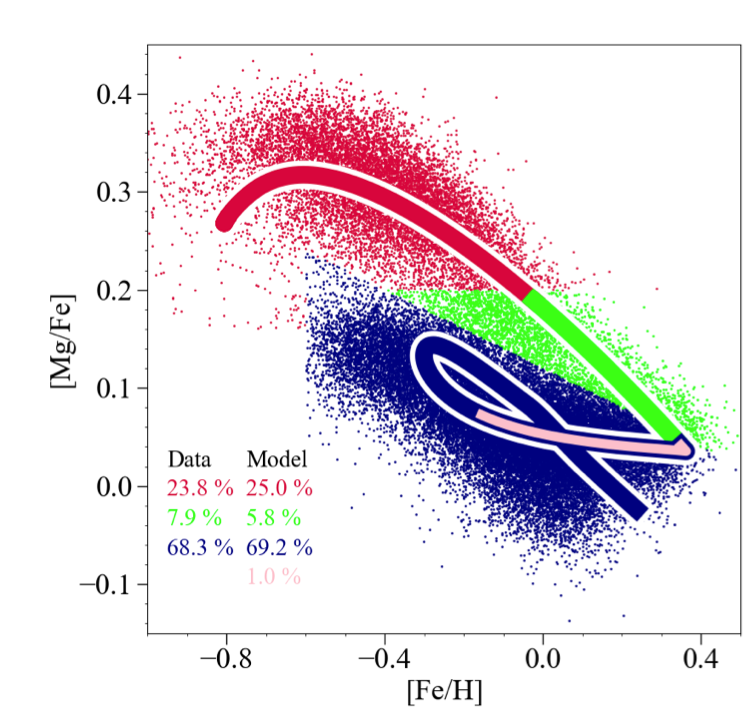}
\caption{The [$\alpha$/Fe] vs. [Fe/H] plot in the solar neighborhood.  The high-$\alpha$ stars are indicated in red, the low $\alpha$ ones in blue and the intermediate ones in green, The pink line indicates the stars formed during the star formation gap in between the two disks (1\%).Figure from \citep{spitoni2024}.}
    \label{fig4}
 \end{figure}

\section{Time-delay model and different galaxies}

Here we show how different galaxies present different trends for the [X/Fe] vs. [Fe/H] relation and the application of the time-delay model to the identification of high-redshift galaxies.\\
In Figure 5  we show model predictions for different galactic environments together with observational data.  In particular, in Figure 5 are reported some old data for the bulge of the MW which are close to the theoretical curve for the bulge, while the curve for the Large Magellanic Cloud (LMC),  well fits the data for this galaxy (\citep{hill2000}) and also those relative to Damped Lyman-$\alpha$  (DLA) systems (\citep{vladilo2002}),
which are high redshift objects illuminated by the light of a very far quasar. Clearly, from this comparison one can conclude that DLA systems are probably irregular galaxies (see \citep{matteucci1997}). This means that the plot [$\alpha$/Fe] vs. [Fe/H]  is a very useful tool to infer the type of galaxies showing a given abundance pattern, and therefore can be used to identify high redshift objects. 
In fact, the [$\alpha$/Fe] vs. [Fe/H] relation is different for each galaxy type: this is due to the different age-[Fe/H] relations deriving from different SFHs in galaxies of different morphological type. The plot in Figure 5 can explain at the same time the data of the MW, the LMC and the bulge (spheroid), and suggests that DLAs are indeed irregular galaxies.
Therefore, by means of this plot we can try to identify also other types of galaxies observed at high redshift, whose morphology is unknown.\\

\begin{figure}
        \includegraphics[width=0.8\linewidth]{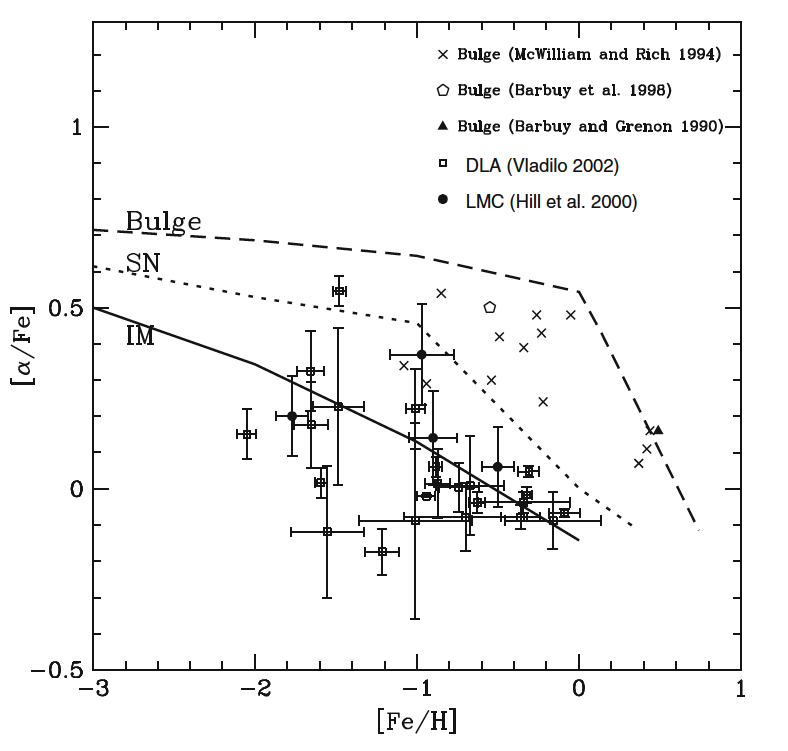}
        \caption{The [$\alpha$/Fe] vs. [Fe/H] for three different galactic environments: the Galacticc bulge, the solar vicinity and a Magellanic irregular galaxy. For $\alpha$ we intend a generic $\alpha$-element such as Mg or O. The data reported in the Figure are for the Galactic bulge, LMC and DLAs, as indicated in the figure.        
        \label{fig5}}
    \end{figure}

\subsection{Lyman-break galaxies are ellipticals}

In Figure 6 we show the estimated SFR for a Lyman-break galaxy at high redshift (z=2.7276)(MS1512-cB58), compared to model predictions for  typical elliptical galaxies. In Figure 7 we show the model predictions for [X/Fe] vs. [Fe/H] for three different elements, compared to the data for CB58. The models are characterized by different total stellar masses (indicated in the plot) and assume a very fast SFR followed by a powerful galactic wind energized by stellar feedback. After the development of a galactic wind, all these galaxies are quenched. The comparison between models and data suggests that cB58 is an elliptical galaxy.  
\begin{figure}
        \centering
        \includegraphics[width=0.7\linewidth]{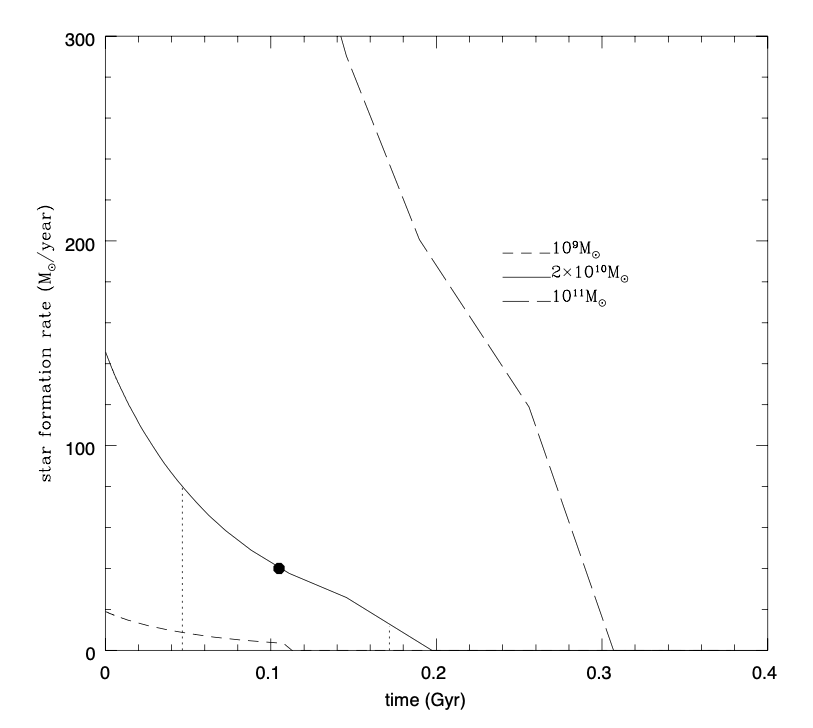}
        \caption{The average star formation rates predicted by the models of \citep{matteucci2002} for ellipticals of different stellar masses, as indicated in the figure, until the onset of the galactic wind inside the optical radius of each galaxy. The dotted lines enclose the range for the possible values of the star formation rate and the full dot represents the value adopted by \citep{pettini2001}.From the comparison between observed and predicted star formation rate one can infer a mass of  $2 \cdot 10^{10}M_{\odot}$ and  an age of $\sim 100 \pm {60}$ Myr for MS 1512-cB58. }       
       
        \label{fig6}
    \end{figure}

\begin{figure}
        \centering
        \includegraphics[width=0.9\linewidth]{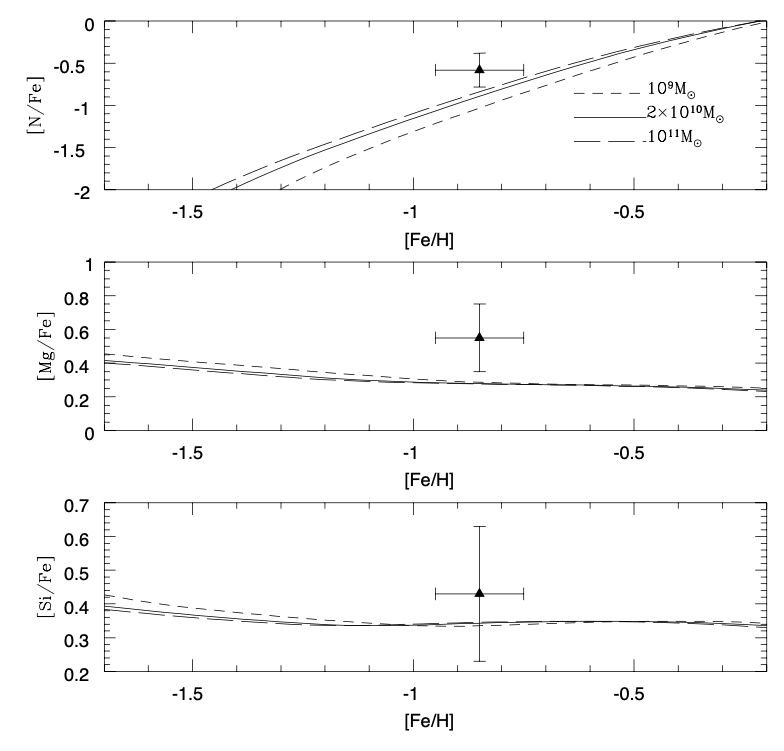}
        \caption{The [X/Fe] vs. [Fe/H] for three different chemical elements: N, Mg and Si. The models refer to spheroids with different stellar masses, as indicated in the plot. The observational data are for the Lyman-break galaxy MS1512-cB58 observed by \citep{pettini2001}. Figure from \citep{matteucci2002}.}           
        \label{fig7}
    \end{figure}

The model predictions, reported in Figure 6,  clearly suggest that to reproduce the observationally estimated SFR, cB58 should have a mass of  $2\cdot 10^{10}M_{\odot}$ and an age of $100 \pm 60$ Myr. Other independent estimates of the age indicate lower ages of 20-35 Myr, but they are affected by the assumed value of the Hubble constant, whereas the chemical age is not. So it was concluded that this Lyman-break galaxy is compatible with a proto-elliptical with active star formation, on the basis of its SFR and abundance patterns reported in Figure 7.
In particular, in Figure 7 the model predictions for ellipticals of different stellar mass show that cB58, taking into account observational errors, is better fitted by models with masses in the range $2 \cdot 10^{10}-10^{11} M_{\odot}$. The observed point relative to [Mg/Fe] is a bit too high compared to the model predictions, but this is a common result for all galaxies, since the literature stellar yields of Mg are  systematically lower than what it would be requested even to fit the Solar Mg abundance, and therefore it is a problem of stellar nucleosynthesis (see \cite{matteucci2021} for a review).

 \subsection{How to identify the hosts of GRBs}       
The hosts of GRBs can be found either locally or at high redshift.
Long-GRBs (LGRB) are thought to be associated to SN explosions, in particular to SNe Ib and Ic which originate from very massive stars. In the work of \citep{grieco2014}, they  applied the time-delay model to identify the hosts of high-redshift LGRB. They compared the abundance patterns predicted for galaxies of different morphological type (spirals, ellipticals and irregulars) with data measured in the galaxies hosting LGRBs. The models included the evolution of dust which influences mostly the predicted Fe abundance.

 \begin{figure}[h]
 \centering
 \includegraphics[width=1.0\linewidth]{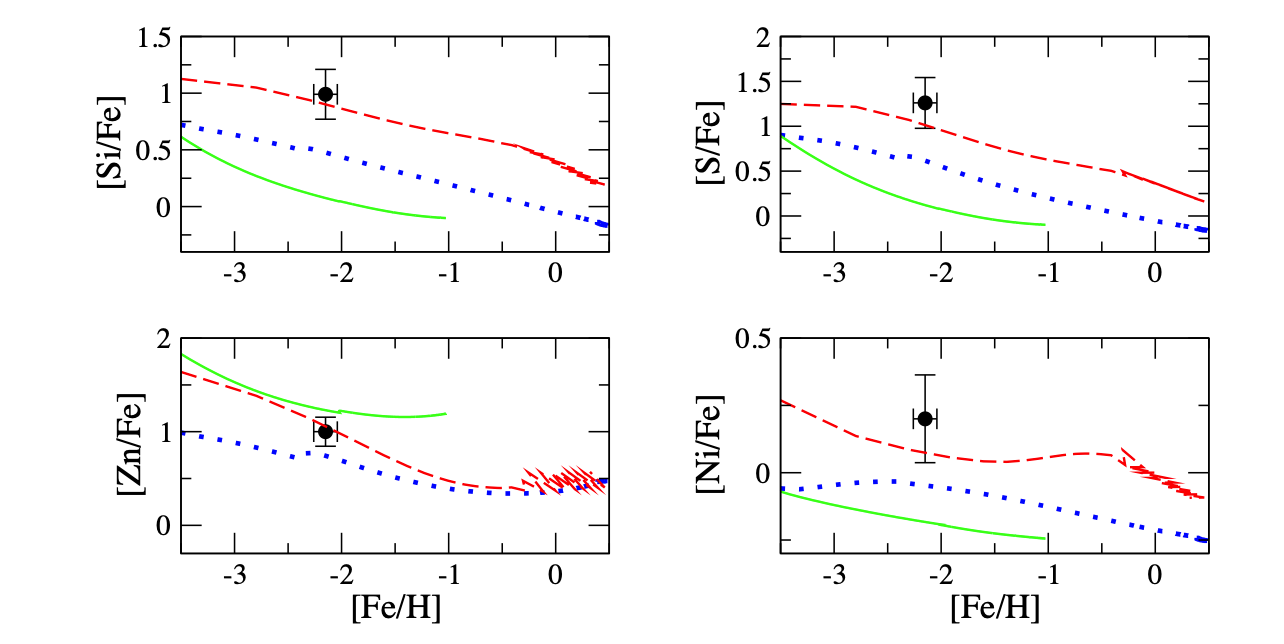}
 \caption{The [X/Fe] vs. [Fe/H] relations for different elements and different galaxies. The dashed red lines refer to a typical elliptical galaxy, the dotted blue lines to a typical spiral like the MW and the continuous green lines to a typical irregular galaxy. Te data with error bars refer to the GRB120815 observed by \citep{kruhler2013}.}
\label{fig8}
\end{figure}

In Figure 8 we show an example of this technique: here model predictions for galaxies of different morphological type (elliptical, spiral and irregular) are compared to data for  GRB129815 at redshift z=2.36 (\citep{kruhler2013}). This comparison clearly shows that we are observing a massive proto-elliptical during its active phase of star formation, since the lines referring to an elliptical of $10^{11}M_{\odot}$ can best fit the data.

\section{Discussion and conclusions}

In this paper, we have shown  how the time-delay model paradigm allows us to reconstruct the history of star formation and assembly of galaxies of different morphological type. Important constraints on stellar nucleosynthesis can be found by comparing model results with observations. The nucleosynthesis of some elements seem now well understood, such as that of $\alpha$-elements (O, Mg, Si, Ca) which are produced mainly by massive stars ($M>8M_{\odot}$) exploding as CC-SNe and some of them are also partly produced by Type Ia SNe, for example Si and Ca. We know that Fe is mainly produced by Type Ia SNe (70\% with an IMF suitable for the solar vicinity) and partly by CC-SNe. Other elements such as $^{12}C$ are mainly produced by massive stars, like the other $\alpha$-elements, and partly by LIMS. This is clear now because of the observed overabundance of C relative to Fe in low metallicity solar vicinity stars. The element $^{14}N$ is mainly produced by LIMS and partly by massive stars and is both a secondary and primary element. It is mainly secondary in LIMS while it is primary in massive stars: this was understood by trying to reproduce the [N/Fe] ratio in halo stars which is much higher than expected if N were produced as a secondary element in massive stars (see \cite{matteucci1986}).
The Fe-peak elements should be mainly formed in Type Ia SN explosions although some of them seem to show a different behavior than expected (Sc for example). For neutron rich elements the problem is to establish the percentages of s- and r- processes that are responsible for their formation. Also in this case, the time-delay model is very useful. We have presented here the evolution of Eu that fortunately is almost entirely a r-process element. By looking at the [Eu/Fe] ratios in the solar neighborhood, we have understood that Eu should be formed by fast stellar producers at early times, such as CC-SNe, merging neutron star systems coalescing on very short times or a higher fraction of these systems at early times. The gravitational event GW170817, which is also the kilonova AT 2017gfo, has proven that such systems are indeed producing very heavy elements. \\
Concerning external galaxies, their abundance patterns have suggested that ellipticals and bulges should have formed very quickly since their [$\alpha$/Fe] ratios are high for their metal content. The contrary happens in irregular galaxies which suffer a mild SFR coupled with ongoing galactic winds that further decrease the gas content and therefore the SFR. In this case, the [$\alpha$/Fe] ratios show a steep decrease with increasing metallicity. In other words, the knee occurring in the [$\alpha$/Fe] vs. [Fe/H] relation varies in galaxies of different morphological type and this is due to the different SFHs.\\
Finally, we have shown that abundance ratios and time-delay model help in identifying high redshift galaxies, whose morphological type is not visible. By means of the above quoted plot, we can suggest that Lyman-break galaxies are ellipticals  and DLAs are irregulars, and we can identify the hosts of LGRBs and this comparison has shown that at high redshift hosts can be galaxies of any type as long as they have active star formation, since the SNe Ib,c which are associated to LGRBs, are connected to short living very massive stars (see also \citep{palla2020}).

\section{Acknowledgements}
F. Matteucci thanks the Italian National Institute for Astronomy (I.N.A.F.) for the 1.05.12.06.05 Theory Grant - Galactic archae-
ology with radioactive and stable nuclei. F. Matteucci also thanks I.N.A.F. for the  1.05.24.07.02 Mini Grant - LEGARE "Linking the chemical Evolution of Galactic discs AcRoss diversE scales: from the thin disc to the nuclear stellar disc" (PI E. Spitoni). F. Matteucci thanks two anonymous referees  who improved the papers with their suggestions.

\bibliography{sn-bibliography}

\end{document}